\documentclass[prd,twocolumn,superscriptaddress,nofootinbib,showpacs]{revtex4}

\usepackage{graphicx,amsmath,amsfonts,amssymb,revsymb,dcolumn,epsfig,bm}

\def\be{\begin{equation}}
\def\ee{\end{equation}}
\def\bea{\begin{eqnarray}}
\def\eea{\end{eqnarray}}

\begin{document}

\title{Limits of the circles-in-the-sky searches  in the determination of cosmic topology of nearly flat
 universes}

\author{G.I. Gomero}
\affiliation{Universidade Estadual de Santa Cruz -- UESC,
Rodovia Ilh\'eus -- Itabuna km 16,  \\
45650-000 Ilh\'eus -- BA, Brazil}

\author{B. Mota}
\affiliation{Instituto de F\'isica, Universidade Federal do Rio de Janeiro ,
Av. Athos da Silveira Ramos 149,  \\
Centro de Tecnologia –- Bloco A - Cidade Universitária  \\
21941-972 Rio de Janeiro -- RJ, Brazil}

\author{M.J. Rebou\c{c}as}
\affiliation{Centro Brasileiro de Pesquisas F\'{\i}sicas,
Rua Dr.\ Xavier Sigaud 150, \\
22290-180 Rio de Janeiro -- RJ, Brazil}

\date{\today}

\begin{abstract}
An important observable signature of  a detectable nontrivial spatial topology
of the Universe is the presence in the cosmic microwave background sky of pairs
of matching circles with the same distributions of temperature fluctuations
--- the so-called circles-in-the-sky.
Most of the recent attempts to find these circles, including the ones undertaken
by the Planck Collaboration, were restricted to antipodal or nearly antipodal circles with
radii $\lambda \geq 15^\circ$.
In the most general search, pairs of circles with deviation from antipodality angles
$0^\circ \leq \theta \leq 169^\circ$ and radii $10^\circ \leq \lambda \leq 90^\circ$
were investigated. No statistically significant pairs of matching circles were
found in the searches so far undertaken.
Assuming that the negative result of general search can be confirmed through analysis
made with data from Planck and future cosmic microwave background (CMB) experiments,
we examine the question
as to whether there are nearly flat universes with compact topology, satisfying Planck
constraints on cosmological parameters, that would give rise to circles-in-the-sky whose
observable parameters  $\lambda$ and $\theta$ fall outside the parameter ranges
covered by this general search.
We derive the expressions for the deviation from antipodality and for the radius of
the circles associated to a pair of elements ($\gamma\,$,$\gamma^{-1}$) of the holonomy group
$\Gamma$ which define the spatial section of any positively curved universe
with a nontrivial compact topology. We show that there is a
critical position that maximizes the deviation from antipodality, and prove that
no matter how  nearly flat the Universe is, it can always have  
a nontrivial spatial topology that gives rise to circles whose deviation from antipodality
$\theta$ is larger than $169^\circ$, and whose radii of the circles $\lambda$
are smaller than $10^\circ$ for some observers's positions. This makes it apparent that
slightly positively curved nearly flat universes with cosmological parameters within
Planck bounds can be endowed with a nontrivial spatial topology
with values of the observable parameters  $\lambda$ and $\theta$  outside the ranges
covered by the  searches for circles carried out so far with either WMAP or Planck data.
Thus, these circles-in-the-sky searches carried out so far are not sufficient to exclude
the possibility of a universe with a detectable nontrivial cosmic topology.
We present concrete examples of lens spaces universes whose associated
circles have both, or at least one value of the observable parameters ($\lambda$, $\theta$)
outside the ranges covered by these searches.
We also present a brief discussion of the implications of our results in view of
unavoidable practical limits of the circles-in-the-sky method.
\end{abstract}


\maketitle

\section{Introduction} \label{S:Intro}

Two fundamental problems regarding the Friedmann-Lema\^{\i}tre-Robertson-Walker
(FLRW) approach to cosmological modeling concern the spatial geometry and topology.
Regarding the geometry, recent high-precision cosmic microwave background (CMB)
data from Planck have provided strong evidence that the universe is  nearly 
flat with  $|\Omega_k| < 0.005$~\cite{Planck-2015-XIII}, which is consistent with
standard inflationary predictions that curvature should be unobservably small
today. Concerning the topology, despite our present-day inability to predict it
from a fundamental theory, one should be able to probe it 
through CMB observations (see, e.g., the reviews~\cite{CosmTopReviews}).

An observable signature of a detectable nontrivial spatial topology is the
presence in the CMB sky of pairs of matching correlated circles with equal
distributions (up to a phase) of temperature fluctuations --- patterns of hot
and cold spots that match around the so-called circles-in-the-sky~\cite{CSS1998}.

Each such pair of circles on the CMB sphere can be specified as a point in
$6$-dimensional parameter space, namely the center of each circle of the
pair (four parameters), the angular radius of both circles (one parameter),
and the relative phase between them (one parameter).
Since such a general search for pairs of circles is very costly in computer
time, most of recent searches, including the ones undertaken by the Planck
team~\cite{Planck-2013-XXVI,Planck-2015-XVIII}, were restricted to back-to-back
circle pairs%
\footnote{This refers to pair of circles whose centers are antipodal points
on the CMB sphere, which are also known as antipodal circles-in-the-sky.}%
~\cite{Cornish-etal-03,Roukema-etal-04,Aurich-etal-05,Aurich-etal-06,Key-et-al-07,%
Bielewicz-Banday-11,Aurich-Lustig-13,Bielewicz-etal-12}
or nearly antipodal circles~\cite{Cornish-etal-03}. No
pairs of matching circles were found through these searches.
This negative result, along with the fact that in a very nearly flat
($|\Omega_k| \lesssim10^{-5}$) compact universe the deviation from
antipodality is small for most observers~\cite{Mota-etal-04,Mota-etal-08},
has been taken to be sufficient to exclude a detectable nontrivial topology
for most observers~\cite{Mota-etal-08}.
However, since the deviation from antipodality in compact orientable {\it exactly}
flat ($|\Omega_k|=0$) universes can be larger than $10$ degrees~\cite{Mota-etal-10},
if the Universe is in fact flat then these restricted searches  for antipodal
or nearly antipodal circles would not be sufficient to rule out the
possibility of a nontrivial flat topology~\cite{Mota-etal-10}.

Although the  $6$-dimensional parameter space can be used 
to statistically extract potential matching circles from CMB maps, it is not
well adapted to study the relation between observable parameters of the circles
and the spatial topology. In fact, while the circle radii and their relative phases
are observable parameters directly linked to the topology, the
positions of the circle centers depend on the choice of the coordinates
and are related to the topology through
the separation angle of circle centers $\Theta$, or equiva\-lently its
supplement $\theta$, which gives the deviation from antipodality.  
Thus,  a more convenient set of parameters for studying
the interrelations between matching circles and the spatial topology,
includes  these observable parameters directly linked to  
the topology, namely the deviation from antipodiciy, $\theta$, the relative
phase angle, $\phi$,  and the angular radii of the circles $\lambda$.

A more general search for circle pairs that are not back-to-back has been carried
out by Vaudrevange \emph{et al.}~\cite{Vaudrevange-etal-12} using
Wilkinson Microwave Anisotropy Probe (WMAP) $7$-years
data~\cite{WMAP7}. 
No statistically significant pairs of matching circles were found.
They have employed the circles-in-the-sky statistics to search for pairs of circles
with radii $ 10^\circ \leq \lambda \leq 90^\circ$ and integer separation angles
of the circle centers $11^\circ \leq \Theta \leq 180^\circ$, extending
the existing lower bounds on these parameters so as to encompass a wider range
of possible topologies. Thus, for example, the negative result of this search, if
confirmed, along with the maximal values of the deviation from antipodality,
$\theta_{\text{max}}$, associated to the shortest geodesics in
multiply-connected orientable flat  manifolds~\cite{Mota-etal-10},
are sufficient to rule out the possibility of a detectable nontrivial orientable
flat cosmic topology whose associated circle radii, $\lambda$, are greater
than $10$ degrees.%
\footnote{In line with the usage in the literature, by the topology of the Universe,
or simply cosmic topology, we mean the topology of its spatial sections.}

Assuming that the negative result of the general
search of Ref.~\cite{Vaudrevange-etal-12} can be confirmed through a similar
analysis made with data from Planck and future CMB experiments,
an important remaining question that naturally arises here is whether there
still are  nearly flat, but not exactly flat, universes with compact topology
that would give rise to circles-in-the-sky whose observable parameters
$\lambda$ and $\theta$ would fall outside the parameter ranges
covered by this more general search.

Our primary objective in this paper is to address this question by considering
nearly flat universes whose spatial section $M$ is a slightly positively curved
space ($\Omega_k \lesssim  0$), whose cosmological parameters are within the
bounds determined by Planck data~\cite{Planck-2015-XIII},
and endowed with a spherical orientable nontrivial topology.
To this end, we first derive the analytic expressions for the deviation
from antipodality and for the radius of the circles of any pair that arises from
a general pair of elements ($\gamma\,$,$\gamma^{-1}$) of the holonomy group 
$\Gamma$ used to form any quotient multiply-connected spherical spaces
$\mathbb{S}^3/\Gamma$.
Second,  for an arbitrary pair ($\gamma\,$,$\gamma^{-1}$) of holonomies
we derive an expression that gives the observer's position in which the deviation
from antipodality attains its maximum.
Third, we then show that no matter how  nearly flat the
Universe is it can always have a nontrivial topology that gives rise to an
observable pair of matching circles whose deviation from antipodality $\theta$ is
larger than $169^\circ$ (or $\Theta < 11^\circ$)  and with radii $\lambda$ smaller
than $10^\circ$ for some observers. Therefore,  with these observable parameters outside
ranges covered by this general search~\cite{Vaudrevange-etal-12}, 
making clear that the circles-in-the-sky searches already undertaken
are not sufficient to exclude the possibility of a detectable
nontrivial topology for the Universe.
Finally, by taking into account the recent bounds on the cosmological parameters by
the Planck Collaboration~\cite{Planck-2015-XIII}, which constrain the distance
to the last scattering surface  $\chi_{\rm{LSS}}^{}$, 
we concretely show examples of a number of lens spaces universes
whose associated circles are such that the value of at least one of the observable
parameters ($\lambda$, $\theta$), or both, falls outside the range covered by
the searches so far undertaken. These examples make apparent that it is possible to
have a universe with a detectable nontrivial cosmic topology that respects Planck constraints
on the cosmological parameters, and have not been excluded by the searches for
circle-in-the-sky carried out so far.

The structure of the paper is as follows.  In Section~\ref{Sec2} we give a
brief account of the  prerequisites necessary for the following sections.
In Section~\ref{S:ThetaLambda} we derive the expression for deviation from
antipodality and for the radius of the circles associated to a pair of elements
($\gamma\,$,$\gamma^{-1}$) $\in \Gamma$ used to form the spatial sections of
multiply-connected universes $\mathbb{S}^3/\Gamma$.
In Section~\ref{S:MaxTheta} we show that no matter how small is the value
for $\chi_{\rm{LSS}}^{}$ (in unit of curvature radius) there are always
holonomies that give rise to observable circles-in-the-sky for which
the maximal deviation from antipodality is detectable  at some observer's 
position, and  it can be made as close to $\pi$ as we require.
In Section~\ref{S:Examples} we construct examples of universes with
lens-space spatial topology whose values of cosmological parameters
respect the bounds of Planck Collaboration~\cite{Planck-2015-XIII},
and for which both or at least one of the observable parameters of
the circles ($\lambda$, $\theta$) fall outside the range covered by
the searches so far undertaken.
In Sec.~\ref{S:Discussion} we present our final remarks and conclusions,
and also briefly discuss the implications of our results in view of
unavoidable practical limits of the circles-in-the-sky method.

\section{Preliminaries} \label{Sec2}   

We begin by recalling the basic cosmological setting of this work. In the context of
general relativity, a fundamental assumption in standard cosmological modelling is
that, on large scales, the Universe is described by a $4$-dimensional manifold
$\mathcal{M} = \mathbb{R}\times M$ endowed with the spatially homogeneous and
isotropic FLRW spacetime metric 
\begin{equation}
\label{FLRW1}
ds^2 = -c^2dt^2 + a^2 (t) \left [ d \chi^2 + f^2(\chi) (d\theta^2 +
\sin^2 \theta  d\phi^2) \right ]\,,
\end{equation}
where $t$ is the cosmic time, $a(t)$ is the scale factor
and $f(\chi) = (\chi, \sin\chi,\sinh\chi)$ depending on the
sign of the constant spatial curvature $k=(0, 1, -1)$.

The  sections $M$ are often assumed to be the simply-connected
$3$-dimensional manifolds: Euclidean $\mathbb{E}^3$, spherical $\mathbb{S}^3$,
or hyperbolic $\mathbb{H}^3$. However, they can also be multiply-connected quotient
$3$-manifolds, which are quotient spaces $M=\widetilde{M}/\Gamma$, where the
covering space $\widetilde{M}$ is the corresponding simply-connected constant
curvature covering manifolds $\mathbb{E}^{3}$, $\mathbb{S}^{3}$ or
$\mathbb{H}^{3}$, and  $\Gamma$ is a discrete and fixed point-free group of
isometries of $\widetilde{M}$ called the covering or holonomy group~\cite{Thurston}.
A generic element of the group $\Gamma$ is denoted by $\gamma$
and called holonomy or simply isometry.

Since we are concerned with nearly flat Universe whose spatial section is a
slightly positively curved space ($\Omega_k \lesssim  0$) allowed by Planck
constraints on cosmological parameter~\cite{Planck-2015-XIII}), in the next
sections we consider that the spatial section of the Universe can be modeled
by any spherical orientable manifold with a nontrivial
topology of the form $M= \mathbb{S}^{3}/\Gamma$.

Regarding the dynamics of the Universe we assume that 
it is given  by a $\Lambda$CDM model, whose matter content is
described by dust with density $\rho_m$, plus a cosmological constant $\Lambda$.
The Friedmann equation can then be written in the form
\begin{equation}   \label{Fried}
a^2 = \frac{k c^2}{ H^2 \left( \Omega - 1 \right) } \;,
\end{equation}
where $H=\dot{a}/a$ is the Hubble parameter, $\Omega = \Omega_m + \Omega_{\Lambda}$ with
$\Omega_m = 8 \pi G \rho_m /\,3 H^2$, $\Omega_{\Lambda}  \equiv 8 \pi G \rho_{\Lambda} /\,3 H^2
= \Lambda /\,3 H^2$, and $G$ is the Newton's constant.

We recall that in the $\Lambda$CDM setting one has that for a slightly positively curved
nearly flat Universe the distance to the last scattering surface  $\chi_{\rm{LSS}}^{}$
in units of the curvature radius is given by%
\footnote{Hereafter we express distances in units of the curvature
radius $a_0= |k|\, H_0^{-1} |\Omega_0-1|^{-1/2}$, and measure angles in radians.}
\begin{equation} \label{Chi_LSS}
\chi_{\rm{LSS}}^{} = \sqrt{|\Omega_{k 0}|}
   \int_1^{1+z_{\rm{LSS}}^{}} \!\!\left[ x^3 \Omega_{m0} +
    x^2 \Omega_{k 0} + \Omega_{\Lambda 0} \right]^{-1/2} dx \,,
\end{equation}
where $\Omega_k = 1-\Omega$ and the subscript $0$ denotes evaluation at
present time. Now, taking into account the recent bounds on the cosmological
parameters by the Planck Collaboration~\cite{Planck-2015-XIII} from Eq.~(\ref{Chi_LSS})
one has $\chi_{\rm{LSS}}^{}= 0.038\,$. We will use this value in the next sections
to construct examples of nearly flat positively curved Universes with
detectable nontrivial topology.

Another important ingredient we shall need in the following sections is
the formulation of the conditions for detectability of cosmic topology.
These conditions were studied for classes of hyperbolic and spherical manifolds
as functions of the cosmological parameters in  Refs.~\cite{TopDetec1,TopDetec2},
and extended to the case of \emph{generic} manifolds in the \emph{inflationary
limit} in Ref.~\cite{Mota-etal-08} (see also Refs.~\cite{Mota-etal-04}).%
\footnote{The question of whether the detection of a non-trivial cosmic
topology can be used to set constraints on cosmological density parameters
has been studied for particular topologies in Refs.~\cite{InverseProb}
(see also the related Ref.~\cite{Mota-etal-11}).}
For this article, we only need to know that a way to study the detectability
in cosmic topology is through the lengths of its closed geodesics as follows.
The length of the closed geodesic generated by $\gamma \in \Gamma $ passing through
a point $\mathbf{u} \in M$ is given by the distance between $\mathbf{u}$ and its image
$\gamma \mathbf{u}$, i.e. by the distance function $d(\mathbf{u},\gamma \mathbf{u})$
in the covering space. This allows the definition of the local injectivity radius
$r_{inj}(\mathbf{u})$ which is half the length of the smallest closed geodesic
passing through the point $\mathbf{u}$. A necessary condition for detectability
of cosmic topology is then given by
\begin{equation}
r_{inj}(\mathbf{u}) < \chi_{\rm{LSS}} \;, \label{detect-cond}
\end{equation}
where $\chi_{\rm{LSS}}$ is the comoving distance evaluated at $z_{\rm{LSS}}^{}$.
For universes whose nontrivial topology are globally homogeneous $r_{inj}(\mathbf{u})$
is position-independent. However, for universes with globally inhomogeneous topology,
the length of the smallest closed geodesic depends on the position $\mathbf{u}$.
Therefore, to determine sufficient conditions for detectability that hold for all
observers we have to use the global (constant) injectivity radius $r_{inj} \equiv
\inf_{\mathbf{u} \in M}\, r_{inj}(\mathbf{u})$, which is the radius of the smallest
sphere inscribable in $M$.

To perform the calculations of the next sections we need the form of
the holonomy transformations $\gamma$ as $4 \times 4$ matrices in $SO(4)$. We
recall that the enumeration of all finite subgroups $\Gamma \in SO(4)$ can be
made in terms of the much simpler enumeration of finite subgroups of $SO(3)$
(for a detailed account on this point we refer the readers to Ref.~\cite{Gaussmann-atal-2001}).
For the connection between $SO(4)$ and $SO(3)$ one uses quaternions, which are
a generalization of the familiar complex numbers with three imaginary unities
$\mathbf{i}$, $\mathbf{j}$ and $\mathbf{k}$ satisfying the non-commutative
multiplication rule
\begin{eqnarray}
\mathbf{i}^2&=&\mathbf{j}^2=\mathbf{k}^2=-1\,, \quad 
\mathbf{i}\mathbf{j}=-\mathbf{j}\mathbf{i}=\mathbf{k}\,, \\ 
\mathbf{j}\mathbf{k}&=&-\mathbf{k}\mathbf{j}=\mathbf{i}\,, \qquad \quad\; 
\mathbf{k}\mathbf{i}=-\mathbf{i}\mathbf{k}=\mathbf{j}\,.
\end{eqnarray}
Given a quaternion $\mathbf{q}=a\mathbf{1} + b\mathbf{i}+c\mathbf{j}+d\,\mathbf{k}$
with $(a,b,c,d)\in \mathbb{R}^4$, one defines the conjugate quaternion as
$\mathbf{q}^*=a\mathbf{1} - b\mathbf{i} - c\mathbf{j} - d\mathbf{k}$ and the
norm of $\mathbf{q}$ by $ |\mathbf{q}|= \sqrt{\mathbf{q}\mathbf{q}^*}=
\sqrt{a^2+b^2+c^2+d^2}$. Quaternions with norm $1$ are called unit quaternions.
For a unit quaternion one has an inverse  given by $\mathbf{q}^{-1} = \mathbf{q}^{*} $.
Geometrically, the set of all quaternions $\mathbf{q}=a\mathbf{1} + b\mathbf{i}+c\mathbf{j}
+d\,\mathbf{k}$ defines the Euclidean space $\mathbb{E}^4$, while the $3$-sphere
$\mathbb{S}^3$ can be identified with the multiplicative group of unit quaternions,
i.e.
\begin{equation}  
\mathbb{S}^3 = \{\mathbf{q} \in {\mathbf{H}} \; / \; |\mathbf{q}| = 1\} \,,
\end{equation}
where $\mathbf{H}$ denotes the associative quaternion algebra over the real
numbers.

\section{Deviation from antipodality and  radii of circles}
\label{S:ThetaLambda}

In this section we  derive the analytic expressions for the deviation
from antipodality and for the radius of the circles of any pair that arises from an
arbitrary pair of elements ($\gamma\,$,$\gamma^{-1}$) of the holonomy group
$\Gamma$ used to form any quotient multiply-connected spherical spaces
$\mathbb{S}^3/\Gamma$.

\subsection{Deviation from antipodality}

Without loss of generality, by a convenient choice of basis an arbitrary
isometry $\gamma \in \Gamma$ can be written in the
form~\cite{Gaussmann-atal-2001}
\begin{equation}
\label{Eq:Isometry}
\gamma = \left[\begin{array}{cccc}
                  \cos \alpha &      0     &      0      & -\sin \alpha \\
                        0     & \cos \beta & -\sin \beta &       0      \\
                        0     & \sin \beta &  \cos \beta &       0      \\
                  \sin \alpha &      0     &      0      &  \cos \alpha
               \end{array}\right] \,,
\end{equation}
where  $\alpha$ and $\beta$ are parameters used to define a generic isometry
of $\mathbb{S}^3$.
We recall that when the distance between a point $\mathbf{p}\in \mathbb{S}^3$ and
its image $\gamma \mathbf{p} \in \mathbb{S}^3 $ is independent of   
$\mathbf{p}$ the isometry $\gamma$ is called Clifford translation, which comes
about if, and only if, $\alpha = \beta$. 

For detectable topologies, the action of each pair of elements ($\gamma\,$,$\gamma^{-1}$)
of the group $\Gamma$ may generate one pair of matching circles in the CMB maps
when the surface of last scattering  (LSS) intersects its two images under the
action of  $\gamma\,$ and $\gamma^{-1}$. When $\gamma\,$ is a Clifford translation
the pair ($\gamma\,$,$\gamma^{-1}$) gives rise to a pair of antipodal matching
circles. As we are interested in both back-to-back and non-antipodal circles in
nearly flat spherical universes with a non-trivial topology, we will focus on  the
general case $\alpha \neq \beta$, and also discuss specific  instances of Clifford
translations. To calculate the distance between an observer and its $\gamma$-image
we have to take into account the observer's position.%
\footnote{Although the calculations carried out in the following hold for an
arbitrary pair of holonomies ($\gamma\,$,$\gamma^{-1}$), in dealing with
detectable matching circles we focus on the holonomies $\gamma\,$'s that
generate the shortest closed geodesic passing through the observer's position
$\mathbf{u} \in M$. Thus in these cases we have $d(\mathbf{u},\gamma \mathbf{u}) =
2\,r_{inj}^{}(\mathbf{u})$.}

To have a qualitative understanding of the calculation below it is useful to bear
in mind the following three points. First, the distance between two points
$\mathbf{u}$ and $\mathbf{v}$ in $\mathbb{S}^3$ is just the angle $\mu$ between
them as seen from the origin of $\mathbb{E}^4$, and thus it is computed through the
scalar product in $\mathbb{E}^4$ as $\langle \mathbf{u},\mathbf{v} \rangle= \cos \mu$.
Second, the line of sight of an observer at $\mathbf{u} \in \mathbb{S}^3$ looking at
$\mathbf{v} \in \mathbb{S}^3$ is given by the tangent vector, at $\mathbf{u}$,
to the geodesic joining these two points. Third, given three points $\mathbf{u}$,
$\mathbf{v}_1^{}$ and $\mathbf{v}_2^{}$ in $\mathbb{S}^3$, the angular separation
between $\mathbf{v}_1^{}$ and $\mathbf{v}_2^{}$ seen by an observer at
$\mathbf{u}$ is the angle between the lines of sight of the observer at
$\mathbf{u}$ looking at $\mathbf{v}_1^{}$ and $\mathbf{v}_2^{}$, and thus it is
calculated through the scalar product of the corresponding tangent vectors at
$\mathbf{u}$.

Now, let $\mathbf{w}_1$ and $\mathbf{w}_2{}$ be the centers of a pair of matched
circles. As measured by the observer, these centers are given by their angular
coordinates in the celestial (unit) sphere $\mathbb{S}^2$, so observationally the angle
between these two points is calculated through their scalar product. However,
since theoretical calculations are done in the $3$-sphere modelled as the set
of unit quaternions, in this approach $\mathbf{w}_1^{}$ and $\mathbf{w}_2^{}$
are lines of sight between the observer 
and its images.  So, they are represented
as tangent vectors, at the observer's position, of the geodesics joining the
observer with each of its images.

In Appendix~\ref{Ap:Tangent} we present the detailed calculations showing that
these tangent vectors are given by
\begin{equation}
\label{CircTangentVec}
\mathbf{w}_1^{} = \frac{1}{\sin \mu} (\mathbf{v}_1^{} - \mathbf{u} \cos \mu) \; \mbox{and}\;
\mathbf{w}_2^{} = \frac{1}{\sin \mu} (\mathbf{v}_2^{} - \mathbf{u} \cos \mu) \, ,
\end{equation}
where we have denoted the distance $d(\mathbf{u},\gamma \mathbf{u})$ simply by $\mu$,
for the sake of brevity.  Equation (\ref{CircTangentVec}) makes apparent that
to have these tangent vectors we need to calculate the distance $\mu$.
To compute this distance, which is nothing but $2 r_{inj}^{}(\mathbf{u})$,
we first note that the holonomy $\gamma$ given by
Eq.~(\ref{Eq:Isometry}) consists of two independent rotations, namely a rotation
of an angle $\alpha$ in the $\mathbf{x}_1^{}$--$\mathbf{x}_4^{}$ plane and a
rotation of an angle $\beta$ in the $\mathbf{x}_2^{}$--$\mathbf{x}_3^{}$ plane.
It follows that one can always choose the coordinate axes in $\mathbb{R}^4$
such that the observer is in the $\mathbf{x}_1^{}$--$\mathbf{x}_2^{}$ plane.
In this way, to give the observer's position one needs only of one
parameter, which is its distance to the axis  $\mathbf{x}_1^{}$. 
Denoting this distance by $\rho$ one has that the observer is given by
\begin{equation} \label{Eq:ObsPosition}
\mathbf{u} = \cos \rho \, \mathbf{1} + \sin \rho \, \mathbf{i} \,.
\end{equation}

The distance between the observer at $\mathbf{u}$ and its image
$\mathbf{v}_1^{}= \gamma \mathbf{u}$ is then given by
\begin{eqnarray}
\label{ImageDist}
\cos \mu & = & \langle \mathbf{u} , \gamma \mathbf{u} \rangle \nonumber \\
         & = & \cos \alpha \cos^2 \rho + \cos \beta \sin^2 \rho \,.
\end{eqnarray}
Clearly, the same expression holds for the distance between $\mathbf{u}$ and its
image $\mathbf{v}_2^{}= \gamma^{-1} \mathbf{u}$.
Now, since we are interested in the deviation from antipodality, which is
$\theta = \pi - \Theta$, where $\Theta$ is the angle between $\mathbf{v}_1$ and
$\mathbf{v}_2{}$, a straightforward calculation whose details are given in
Appendix~\ref{Ap:Theta} yields
\begin{eqnarray}
\label{AntipodDist}
\cos \theta & = & - \langle \mathbf{w}_1^{} , \mathbf{w}_2^{} \rangle \nonumber \\
            & = & 1 - \frac{2}{\sin^2\mu} \, (\cos \alpha - \cos \mu)
                   (\cos \mu - \cos \beta)
\end{eqnarray}
for the deviation from antipodality of a pair of circles that arises from an
arbitrary pair of holonomies ($\gamma\,$,$\gamma^{-1}$) of the  group $\Gamma$
used to form any generic quotient multiply-connected spherical spaces
$\mathbb{S}^3/\Gamma$.

It should be noted  that the deviation from antipodality depends on the
position of the observer given by $\rho$, since from Eq.~(\ref{AntipodDist})
one clearly has that $\theta$ depends on $\mu$ (the distance between the observer
and its $\gamma$--images), and from Eq.~(\ref{ImageDist}) $\mu$ depends
on $\rho$. Moreover, it follows from Eq.~(\ref{AntipodDist}) that the matched
circles are antipodal ($\theta = 0$) for $\mu = \alpha$ and $\mu = \beta$;
i.e., when the observer is at any of the two limiting positions $\rho = 0$
or $\rho = \pi/2$. For the observer's position in between these limiting values
the matching correlated circles are not antipodal, and there are positions
for which maximal deviations of antipodality come about.

In Appendix \ref{Ap:CriticalMu} we present detailed calculations
showing that for an arbitrary pair of holonomies ($\gamma\,$,$\gamma^{-1}$), the
critical value $\mu_0^{}$ for which the deviation from antipodality attains
its maximum value is given by
\begin{equation}
\label{CriticalMu}
\cos \mu_0^{} = \frac{1 + \cos (\alpha + \beta)}{\cos \alpha + \cos \beta}\,.
\end{equation}
{}From this equation along with Eq.~(\ref{AntipodDist}) we can derive the
expression of the maximal deviation from antipodality for any
given pair of holonomies, which is given by 
\begin{equation}
\label{MaxTheta}
\theta_{\mbox{\small max}}^{} = |\beta - \alpha| \; .
\end{equation}

Finally,  from Eqs.~(\ref{ImageDist}) and~(\ref{CriticalMu}) one
has  that this maximal deviation from antipodality is detectable by an observer
at a position $\rho_0$ given by
\begin{equation} \label{rho0}
\cos^2 \rho_0^{} = \frac{\sin \beta}{\sin \alpha + \sin \beta}
\end{equation}
whenever the injectivity radius $r_{inj}^{}$ at $\rho_0^{}$ is less than the
distance to the LSS, i.e. whenever the necessary condition for detectability
of cosmic topology $\mu_0^{} < 2\chi_{\rm{LSS}}^{}$ is fulfilled at  $\rho_0^{}$.

\subsection{Angular radius of circles}

To derive the expression for the angular radius of a generic matching circle-in-the-sky
that arises from an arbitrary pair of holonomies ($\gamma\,$,$\gamma^{-1}$),
we begin by recalling that $\chi_{\rm{LSS}}^{}$ denotes the comoving radius of
the last scattering sphere.

To have a sketch of the calculations given in details below,  let
$[\mathbf{u},\mathbf{v}_1^{}]$ be the geodesic segment joining the observer and
its image $\mathbf{v}_1^{}=\gamma \mathbf{u}$, and let $\mathbf{p}$ be the middle point of this
segment. Thus, a point $\mathbf{q}$ in the matching circle generated by the holonomy
$\gamma$ lies in the geodesic plane orthogonal to $[\mathbf{u},\mathbf{v}_1^{}]$
at $\mathbf{p}$, and is distant $\chi_{\rm{LSS}}^{}$ from $\mathbf{u}$ and
$\mathbf{v}_1^{}$. Now, given the segments of geodesics $[\mathbf{u},\mathbf{p}]$
and $[\mathbf{u},\mathbf{q}]$, one can easily find the tangent vectors
$\mathbf{w}_p^{}$ and $\mathbf{w}_q^{}$ to these segments at the observer's
position $\mathbf{u}$. The angular radius $\lambda$ of the matching circle is
then given by the scalar product
\begin{equation}
\label{lambda}
\cos \lambda = \langle \mathbf{w}_p^{}, \mathbf{w}_q^{} \rangle \; .
\end{equation}

To effectively compute the angular radius of matched circles it is convenient
to choose a suitable coordinate system.
In fact, without loss of generality the calculation can be carried out more
easily by choosing a coordinate system such that the observer's position is
$\mathbf{u}=\mathbf{1}$, and thus its image is
\begin{equation}
\mathbf{v}_1^{} = \gamma\mathbf{u} =\cos \mu \, \mathbf{1} + \sin \mu \, \mathbf{i} \, .
\end{equation}
The middle point of the geodesic segment $[\mathbf{u},\mathbf{v}_1^{}]$  is then
\begin{equation}
\mathbf{p} = \cos \frac{\mu}{2} \, \mathbf{1} + \sin \frac{\mu}{2} \, \mathbf{i} \, ,
\end{equation}
and a generic point of the circle associated to the holonomy $\gamma$  has the form
\begin{equation}
\mathbf{q} = t\mathbf{p} + y \mathbf{j} + z \mathbf{k} \, ,
\end{equation}
where $t$, $y$ and $z$ are real numbers subject to the conditions
\begin{equation}
|\mathbf{q}| = 1 \qquad\mbox{and}\qquad \cos \chi_{\rm{LSS}}^{} = \langle
\mathbf{1},\mathbf{q} \rangle \, .
\end{equation}
The tangent vectors at $\mathbf{u}$ to the segments of geodesics
$[\mathbf{u},\mathbf{p}]$ and $[\mathbf{u},\mathbf{q}]$ are then
given, respectively, by
\begin{eqnarray}
\mathbf{w}_p^{} &=& \frac{1}{\sin \frac{\mu}{2}} \left(\mathbf{p} -
\cos \frac{\mu}{2} \, \mathbf{1}\right)\,, \label{wp} \\ 
\mathbf{w}_q^{} &=& \frac{1}{\sin \chi_{\rm{LSS}}^{} } (\mathbf{q} -
\cos \chi_{\rm{LSS}}^{} \, \mathbf{1}) \label{wq} \, .
\end{eqnarray}
Substituting Eqs.~(\ref{wp})~and~(\ref{wq}) into Eq.~(\ref{lambda}),
a simple calculation yields
\begin{equation} \label{CircleRad}
\cos \lambda = \frac{\tan (\,\mu/2\,)}{ \tan \chi_{\rm{LSS}}^{}} \,,  
\end{equation}
for the angular radius of each matching circles generated by
a general pair of holonomies ($\gamma\,$,$\gamma^{-1}$) of the group
$\Gamma$ used to form any multiply-connected spherical space
$\mathbb{S}^3/\Gamma$.

Before proceeding to the next section, it is important to note that both, the deviation
from antipodality and the radii of the circles-in-the sky, depend not only
on the parameters $\alpha$ and $\beta$ that defines  the holonomy $\gamma$,
but also on the position of the observer determined by the parameter $\rho$
through the expression (\ref{ImageDist}). This latter dependence, often 
neglected in the literature,  will be of the utmost importance in the following
sections.

\section{Upper bound on the deviation from antipodality} \label{S:MaxTheta}

In this section we show that, for any observational value of $\chi_{\rm{LSS}}^{}$,
no matter how small it is, there are always holonomies that give rise to
circles-in-the-sky for which the maximal deviation from antipodality
$\theta_{\mbox{\small max}}^{}$ is detectable,
and this deviation can be made as close to $\pi$ as we want.%
\footnote{Concrete examples of nearly flat spherical multiply-connected universe
that illustrate this point are given in Sec.~\ref{S:Examples}.}

A general sketch of the proof is as follows. We begin by defining parameters
$\epsilon$ and $\delta$ by the relations
\begin{equation} \label{epsdelt}
\frac{\beta + \alpha}{2} = \frac{\pi}{2} - \epsilon \qquad\mbox{and}\qquad
\frac{\beta - \alpha}{2} = \frac{\pi}{2} - \delta \; ,
\end{equation}
and rewriting Eqs.~(\ref{CriticalMu}) and~(\ref{rho0}), respectively,  as
\begin{equation}
\label{CriticalMuSin}
\cos \mu_0^{} = \frac{\sin \epsilon}{\sin \delta}
\quad \mbox{and} \quad
\cos^2 \rho_0^{} = \frac{\sin (\delta+\epsilon)}{2\,\sin\delta\,\cos\epsilon} \; .
\end{equation}
We clearly have $\alpha = \delta - \epsilon$ and $\beta = \pi - (\delta +
\epsilon)$, so the maximal deviation from antipodality given by Eq.~(\ref{MaxTheta})
reduces to
\begin{equation} \label{MaxTheta-delta}
\theta_{\mbox{\small max}}^{} = \pi - 2\delta\,,
\end{equation}
which shows that a maximal deviation from antipodality close to $\pi$ requires a
small value of $\delta$.
On the other hand, the condition for detectability of cosmic topology at $\rho_0$
along with the near flatness of the universe is ensured, respectively, by the
inequalities $\mu_0^{} < 2\chi_{\rm{LSS}}^{} \ll 1$.
Together with the first equation~(\ref{CriticalMuSin}), these inequalities
require that $\delta \approx \epsilon$, which in turn leads a very small value
for $\alpha$. Since  $\delta$ has a very small value, so does $\epsilon$, and
thus $\beta$ is very close to $\pi$.

To implement this general reasoning, consider the canonical generator of the lens
space $L(p,q)$, with $p$ prime, given by Eq.~(\ref{Eq:Isometry}) with
\begin{equation} \label{lens-spa}
\alpha = \frac{2\pi}{p} \qquad \mbox{and} \qquad \beta = \frac{2\pi q}{p} \; .
\end{equation}
Equations~(\ref{epsdelt}) and (\ref{lens-spa}) yield
\begin{equation}
\label{Eq:EpsDelta}
\epsilon = \left[\frac{p - 2(q+1)}{p}\right] \frac{\pi}{2}
\;\; \mbox{and} \;\;
\delta = \left[\frac{p - 2(q-1)}{p}\right] \frac{\pi}{2} \,,
\end{equation}
and thus
\begin{equation}
\label{Eq:EpsilonDelta}
\frac{\epsilon}{\delta} = 1 - \frac{4}{p-2(q-1)} \; .
\end{equation}

Given any positive $\delta_0^{} \ll 1$, which defines a maximal deviation
from antipodality threshold $\theta_0^{} \equiv \pi - 2\delta_0^{}$, we will now show
how to choose values for $p$ and $q$ such that $0 < \epsilon < \delta < \delta_0^{}$
and $\epsilon/\delta$ as close to $1$ as we want ($\epsilon/\delta \approx 1$)
in such way that
\begin{equation}
\label{Eq:MuZero}
\cos \mu_0^{} = \frac{\sin \epsilon}{\sin \delta} > \frac{\epsilon}{\delta} >
\cos 2\chi_{LSS}^{} \,.
\end{equation}
Since this equation along with Eq.~(\ref{MaxTheta-delta}) imply that $\mu_0^{} <
2\chi_{\rm{LSS}}^{}$ and $\pi > \theta_{\mbox{\small max}}^{} >\theta_0^{}$,
this amounts to saying that such nearly flat spherical universes could have the topology of
a lens space with detectable maximal deviation from antipodality of matching  
circles as close to $\pi$ as we require.

To insure that $\epsilon/\delta > 0$, from ~(\ref{Eq:EpsilonDelta}) one clearly
has that  $p-2(q-1) > 4$, and thus
\begin{equation}  \label{PermissibleQ}
p > 2q+2 \,.
\end{equation}
Hereafter, for each value of $p$, we shall refer to the values of $q>1$
satisfying this inequality as permissible values.%
\footnote{Here, we do not allow $q=1$ since this corresponds to a Clifford translation,
which gives rise to pairs of antipodal circles, i.e. $\theta=0^\circ$.}
The least impermissible value of $q$ is given by
\begin{equation} \label{Defq0}
q_0^{} = \frac{p-1}{2} \,,
\end{equation}
since for $n=1,2,3, \ldots, q_0^{}-2$, values of $q$ given by
\begin{equation}
\label{Defqn}
q_n^{} = q_0^{} - n
\end{equation}
satisfy Eq.~(\ref{PermissibleQ}) and so are all permissible values.
Since $q_n^{} > 1$ for globally inhomogeneous spaces, from equations~(\ref{Defq0})
and~(\ref{Defqn}) it follows that $p > 2n+3$.

Substituting~(\ref{Defq0}) and~(\ref{Defqn}) into~(\ref{Eq:EpsDelta})
one obtains
\begin{equation} \label{AnglesN}
\epsilon_n^{} = \left(\frac{2n-1}{p}\right) \frac{\pi}{2}
\qquad\mbox{and}\qquad
\delta_n^{} = \left(\frac{2n+3}{p}\right) \frac{\pi}{2} \; ,
\end{equation}
and thus
\begin{equation} \label{eps-n-delt-n}
\frac{\epsilon_n^{}}{\delta_n^{}} = \frac{2n-1}{2n+3} \; .
\end{equation}
Equation~(\ref{eps-n-delt-n}) shows that for a large enough value of $n$ we can
have $\epsilon_n^{}/\delta_n^{} \approx 1$, and the second equation~(\ref{AnglesN})
shows that, additionally, for a large enough value $p$, we can have $\delta_n^{} <
\delta_0^{}$, as we wanted to show.

Finally, we note that in  the construction of the concrete examples of nearly flat
multiply-connected universes we need a more precise estimate for the values of $n$
in terms of $\chi_{\rm{LSS}}^{}\,$. To this end,  from
Eq.~(\ref{eps-n-delt-n}) one has that the second inequality in~(\ref{Eq:MuZero}),
i.e. $\epsilon_n^{}/\delta_n^{} > \cos 2\chi_{\rm{LSS}}^{}\,$, holds if
\begin{equation} \label{Eq:NXLSS}
n > \frac{1 + 3 \cos 2\chi_{LSS}^{}}{2(1 - \cos 2\chi_{LSS}^{})} \; .
\end{equation}

\section{Concrete Examples} \label{S:Examples}  

As a first illustrative example, consider a nearly flat universe with a lens-space
spatial topology, and whose values of cosmological parameters respect the constraints
on the cosmological parameters
determined by Planck Collaboration~\cite{Planck-2015-XIII} so that the distance to the
LSS in units of the curvature radius is $\chi_{\rm{LSS}}^{} = 0.038$ [cf. Eq.~(\ref{Chi_LSS})].
Let us then look for a detectable holonomy that gives rise to circles-in-the-sky with
detectable maximal deviation from antipodality larger than $99\pi/100$, which 
from Eq.~(\ref{MaxTheta-delta}) gives $\delta < \pi/200$.  
Besides, from equation~(\ref{Eq:NXLSS}) one has that $n = 692$ guarantees that
$\epsilon_n^{}/\delta_n^{} > \cos 2\chi_{\rm{LSS}}^{}$. Equations~(\ref{AnglesN})
in turn give
\begin{equation}
\epsilon_{692}^{} = \frac{1383\pi}{2p}
\qquad\mbox{and}\qquad
\delta_{692}^{} = \frac{1387\pi}{2p} \,.
\end{equation}
Thus, to assure that $\delta_{692}^{} < \pi/200$ one has $p > 138\,700$.
Since we are looking for $p$ prime (this is not strictly necessary), we take
the next prime greater than this value, i.e. $p= 138\,727$ and obtain from Eq.~(\ref{Defq0})
the least impermissible value of $q_0^{} = 69\,363$, and thus $q_{692}^{} = 68\,671$.
In brief, this shows that in a nearly flat universe with $\chi_{\rm{LSS}}^{} = 0.038$,
a maximal deviation from antipodality larger than $99\pi/100$ is
detectable through the circles associated with the pair of holonomies  
($\gamma, \gamma^{-1}$) defined by  Eq.~(\ref{lens-spa}), i.e. given by the
canonical generator of the lens space $L(138\,727,68\,671)$. In fact,
in this case, one has
\begin{equation}
\theta_{\mbox{\small max}}^{} = \pi - 2\delta_{692}^{} \approx 0.990002\,\pi >
\frac{99\pi}{100} \,.
\end{equation}

In the following, we again take into account the Planck constraints on the
cosmological density parameters~\cite{Planck-2015-XIII} that permit
$\chi_{\rm{LSS}}^{} = 0.038$, and present concrete examples (with smaller values for $p$)
of nearly flat universes with lens-space spatial topology that give rise to detectable
circles-in-the-sky in which either values of the observable parameters $\lambda$ (radius of circles) or
$\theta$ (deviation from antipodality), or even both values, fall outside the ranges covered
by the searches so far undertaken~\cite{Cornish-etal-03,Roukema-etal-04,Aurich-etal-05,%
Aurich-etal-06,Key-et-al-07,Bielewicz-Banday-11,Aurich-Lustig-13,Bielewicz-etal-12,%
Vaudrevange-etal-12}.                

\begin{table}[ht!]
\begin{tabular}{*4{c}}  
\hline  \hline
\ \ \; $L(p,q^\bot)$ \; \ \  & \ \ \; $\mu_0^{}$ \; \ \ & \ \ \; $\lambda$ \; \ \ \\
\hline
\; (24\,907,\,11\,763) \; &  0.075978  &  $1.4^\circ$ \\
(24\,943,\,11\,780) &  0.075923  &  $2.6^\circ$ \\
(24\,967,\,11\,791) &  0.075868  &  $3.4^\circ$ \\
(25\,013,\,11\,813) &  0.075814  &  $4.0^\circ$ \\
(25\,057,\,11\,834) &  0.075759  &  $4.6^\circ$ \\
(25\,087,\,11\,848) &  0.075705  &  $5.1^\circ$ \\
(25\,111,\,11\,859) &  0.075650  &  $5.5^\circ$ \\
(25\,147,\,11\,876) &  0.075596  &  $5.9^\circ$ \\
(25\,183,\,11\,893) &  0.075542  &  $6.3^\circ$ \\
(25\,219,\,11\,910) &  0.075488  &  $6.7^\circ$ \\
\hline\hline
\end{tabular}
\caption{Globally inhomogeneous lens spaces $L(p,q)$ spatial topology of nearly
flat universes with $\chi_{\rm{LSS}}^{} = 0.038$, for which both the maximal deviation
of antipodality, $\theta_{\mbox{\small max}}^{}$, and the radius of the matching
circles, $\lambda$, fall outside the range covered in the searches carried out
with either WMAP or Planck data~\cite{Cornish-etal-03,Roukema-etal-04,Aurich-etal-05,%
Aurich-etal-06,Key-et-al-07,Bielewicz-Banday-11,Aurich-Lustig-13,Bielewicz-etal-12,%
Vaudrevange-etal-12}.
As explained in Appendix~\ref{Ap:Tables},  $q^\bot$ is the least value of $q$
for which $\theta_{\mbox{\small max}}^{} > 170^\circ$, and $\mu_0^{}$ is the critical
value of the distance of the observer and its images for which the deviation from
antipodality attains its maximum. The observer's position
is given by $\rho_0$, which from Eq.~(\ref{ImageDist}) depends on $\mu_0$ for any
given holonomy pair ($\gamma, \gamma^{-1}$).
In  these universes $\theta_{\mbox{\small max}}^{} \approx 170^\circ$ for the
observer at $\rho_0\,$ [cf. Eq.~(\ref{CriticalMuSin})].  \label{Tb:Antipod1} } 
\end{table}

In Table~\ref{Tb:Antipod1} we collect together examples of nearly flat lens-space
universes with $\chi_{\rm{LSS}}^{} = 0.038$ for which both
parameters, the maximal deviation of antipodality, $\theta_{\mbox{\small max}}^{}$, and the
radius of the matching circles, $\lambda$, fall outside the range covered
by the most general search for circles~\cite{Vaudrevange-etal-12}. For full details about
the calculations for the construction of Table~\ref{Tb:Antipod1} we refer the readers
to Appendix~\ref{Ap:Tables}.

Although in all universes of Table~\ref{Tb:Antipod1} both the radii of the circles,
$\lambda$, and the deviation from antipodality,  $\theta$, fall outside the ranges
covered by the searches carried out so far, clearly we only need that one of these
parameters falls outside those ranges to have nearly flat universes
with a nontrivial lens-space topology which has not been ruled out by the searches 
so far undertaken with either WMAP or  Planck data~\cite{Cornish-etal-03,Roukema-etal%
-04,Aurich-etal-05,Aurich-etal-06,Key-et-al-07,Bielewicz-Banday-11,Aurich-Lustig-13,%
Bielewicz-etal-12,Vaudrevange-etal-12}.
In Table~\ref{Tb:Antipod2} we collect examples of nearly flat universes (with
$\chi_{\rm{LSS}}^{} = 0.038$) endowed with lens space spatial topology
with detectable maximal deviation from antipodality
$\theta_{\mbox{\small max}}^{}\gtrsim 90^\circ$,
so within the range covered the most general search~\cite{Vaudrevange-etal-12}.
Nevertheless, with radii $\lambda < 10^\circ$, so outside the range covered by these searches, and
therefore not excluded by all searches so far undertaken.
In Appendix~\ref{Ap:Tables} we give the full details of de calculations made
for the construction of this Table.

\begin{table}[ht!]
\begin{tabular}{*4{c}}  
\hline  \hline
\ \ \; $L(p,q^\bot)$ \; \ \  & \ \ \; $\mu_0^{}$ \; \ \ & \ \ \; $\lambda$ \; \ \ \\
\hline
\; (2\,203,552) \; &  0.075625  &  $5.7^\circ$ \\
(2\,207,\,553) &  0.075556  &  $6.2^\circ$ \\
(2\,213,\,555) &  0.075507  &  $6.5^\circ$ \\
(2\,221,\,557) &  0.075370  &  $7.4^\circ$ \\
(2\,237,\,561) &  0.075100  &  $8.8^\circ$ \\
(2\,239,\,561) &  0.075013  &  $9.3^\circ$ \\
(2\,243,\,562) &  0.074946  &  $9.6^\circ$ \\
(2\,251,\,565) &  0.074917  &  $9.7^\circ$ \\
(2\,267,\,572) &  0.074961  &  $9.5^\circ$ \\
(2\,269,\,572) &  0.074876  &  $9.9^\circ$ \\
\hline\hline
\end{tabular}
\caption{Globally inhomogeneous lens spaces $L(p,q)$ spatial topology of nearly
flat universes with $\chi_{\rm{LSS}}^{} = 0.038$, for which only the
radii of the matching circles, $\lambda$, fall outside the range
covered in the searches so far undertaken. 
As explained in Appendix~\ref{Ap:Tables},  $q^\bot$ is the least
value of $q$ for which $\theta_{\mbox{\small max}}^{} > 90^\circ$, and
$\mu_0^{}$ is the critical value for which the deviation from
antipodality attains its maximum. The observer's position is given
by $\rho_0$ [cf. Eq.~(\ref{CriticalMuSin})], which from Eq.~(\ref{ImageDist})
depends on $\mu_0$ for any given holonomy pair ($\gamma, \gamma^{-1}$).
In  these universes $\theta_{\mbox{\small max}}^{} \gtrsim 90^\circ$
for the observer at $\rho_0\,$. Note that the values of the parameters
$p$ and $q$ are considerably smaller than those in Table~\ref{Tb:Antipod1}. }
\label{Tb:Antipod2}
\end{table}

Given that globally homogenous spaces give rise to antipodal circles ($\theta = 0^\circ$),
an important remaining question that naturally arises here is whether there  are nearly
flat (with $\chi_{\rm{LSS}}^{} = 0.038$) globally homogeneous lens-space universes
that give rise to circles-in-the-sky with radii smaller than $10^\circ$, i.e. outside
the range covered by the most general search so far undertaken~\cite{Vaudrevange-etal-12}.
In the remainder of this section we shall examine this question.

We begin by recalling that in globally homogeneous lens spaces $L(p,q)$ one 
has $q=1$, and that the injectivity radii $ 2\,r_{inj}^{}(\mathbf{u}) \equiv \mu=
d(\mathbf{u},\gamma \mathbf{u}) $ for these spaces are constants and given by
$\mu = 2\pi/p\,$~\cite{TopDetec1}. Hence, equation~(\ref{CircleRad}) reduces to
\begin{equation}
\label{CircleRadHom}
\cos \lambda = \frac{\tan \frac{\pi}{p}}{\tan \chi_{\rm{LSS}}^{}} \; .
\end{equation}
Thus, in order to have a pair of matched circles with radius smaller
than a certain value $\lambda_0^{}$ (say) one must have
\begin{equation}
\label{IneqHomLens1}
p < \frac{\pi}{\arctan (\cos \lambda_0^{} \tan \chi_{LSS}^{})} \,,
\end{equation}
which along with the condition for detectability of the topology,
$\mu < 2 \chi_{LSS}^{}\,$, leads to
\begin{equation}
\label{IneqHomLens2}
p > \frac{\pi}{\chi_{LSS}^{}} \; .
\end{equation}
For $\lambda_0^{} = 10^\circ$ and $\chi_{LSS}^{} = 0.038$ one obtains
\begin{equation}
82.7 < p < 83.95 \,.
\end{equation}
Therefore,  $\chi_{\rm{LSS}}^{} = 0.038$ allows for only one nearly flat globally
homogeneous lens space universe that has not been ruled out by the searches so
far undertaken, namely $L(83,1)\,$. This lens space universe would give rise to
antipodal circles with radius smaller than  $10^\circ$.

The extent to which the examples of this section depend upon the uncertainties in the
determination of the cosmological parameters, which give rise to uncertainties on the
determination of $\chi_{LSS}^{}$, is an important point to be investigated, beyond the
scope of the present article, though.

\section{Concluding remarks} \label{S:Discussion}

The existence in the CMB anisotropy maps of correlated pairs of circles
with the same distribution of temperature fluctuations (up to a phase),
the so-called circles-in-the-sky, is a key prediction for a universe with
a detectable non-trivial cosmic topology.
Detecting such circles, and measuring the relative position of their
centers, angular radii, and relative phase, would allow us to possibly
determine the topology of the spatial section of the Universe.

Most of the recent searches, including the ones undertaken by the Planck Collaboration,
were restricted to circles whose centers are antipodal points on the CMB sphere 
or nearly antipodal  circles. 
In the most general search~\cite{Vaudrevange-etal-12} the circles-in-the-sky
statistics was employed to search for pairs of circles with radii
$ 10^\circ \leq \lambda \leq 90^\circ$ and integer
deviation from antipodality angles $0^\circ \leq \theta \leq 169^\circ$.

In this work, 
we have addressed the question as to whether there are still nearly flat slightly
positively curved universes ($\Omega_k \lesssim  0$), that respect Planck
constraints on the cosmological parameters~\cite{Planck-2015-XIII}, and
have compact spatial topology that would give rise to circles in CMB sphere
whose observable parameters $\lambda$ (radius) and $\theta$ (deviation
from antipodality) fall outside the parameter ranges covered by this
general search.

To answer this question, we have derived the analytic expressions for the deviation
from antipodality and for the radius of the circles that arise from a pair of
elements ($\gamma\,$,$\gamma^{-1}$) of the holonomy group $\Gamma$ which
is used to define the spatial section of any positively curved universe
with a nontrivial compact topology.
Since the deviation from antipodality depends upon the observer's position
$\rho$, we have derived the expression for the critical position $\rho_0$ that
maximizes the deviation from antipodality in a universe with a generic
$\mathbb{S}^3/\Gamma$ spatial topology.
We then have shown that no matter how  nearly flat the Universe is, it can always
have a nontrivial topology whose deviation from antipodality $\theta$ of
the associated  pair of matching circles is larger than $169^\circ$ and
whose radii $\lambda$ are smaller than $10^\circ$ for some observers's
positions. 
This important result makes apparent that slightly positively curved nearly flat
universes can always be endowed with a nontrivial spatial topology that gives rise
to circles with two observable parameters associated to the topology,  $\lambda$
and $\theta$, outside the range covered in the searches carried out so far
with either WMAP and Planck data~\cite{Cornish-etal-03,Roukema-etal-04,Aurich-etal-05,%
Aurich-etal-06,Key-et-al-07,Bielewicz-Banday-11,Aurich-Lustig-13,Bielewicz-etal-12,%
Vaudrevange-etal-12}. This makes clear that the circles-in-the-sky searches already
undertaken are not sufficient to exclude the possibility of a universe with
a detectable nontrivial cosmic topology, and that respects Planck constraints on the
cosmological parameters.

By taking into account the recent constraints on the cosmological parameters by
the Planck Collaboration~\cite{Planck-2015-XIII}, we have also exhibited concrete
examples of lens spaces universes whose associated circles have both, or at least
one value of the observable parameters $\lambda$ and $\theta$ outside the
ranges covered by the searches so far undertaken. The existence of such concrete
examples with very small radii of the circles, which are statistically difficult
or even impossible to detect, has been conjectured in the recent
Ref.~\cite{Luminet-2016}.

Although only the limited range for deviation from antipodality  $0^\circ \leq \theta \leq 169^\circ$
has been explored until now, the future searches for circles can in principle
cover the whole range of values $0^\circ \leq \theta \lesssim 180^\circ$.  
However, regarding the radii of the circles, two facts should be taken into account.
First, the peak amplitude in the circles statistic, $S^{\mbox{\small max}}$, decreases with
the radius of the circles due primarily to the Doppler term, which is increasing anticorrelated
for circles with radius smaller than $45^\circ\,$~\cite{Cornish-etal-03}. Second,
the intersection of the peaks of the circle statistics with the false detection threshold
line defines the smallest circle radius that one could expect to detect.
Since the false positive line is higher
for circles with smaller radius, in practice there is a minimum value for the
radius of the circles we could expect to detect. The smaller value achieved in the search
so far  is around $\lambda \simeq 10^\circ$ with an estimated value  $\lambda \simeq 5^\circ$
for future searches using polarized CMB maps~\cite{Bielewicz-etal-12}. The first
five entries in Table~\ref{Tb:Antipod1} are examples of nearly flat universes,
respecting Planck constraints on the cosmological parameters, that give rise
to matching circles with radii smaller than this minimum estimated value.
For these lens spaces universes the unavoidable practical limits for small radii
of the circles-in-sky method, due to the combination of the falling of the peaks
amplitude with the increasing of the false detection threshold,
are reached. Thus, the spatial topology of these universes cannot be unveiled  
through the circle-in-the-sky method.


\begin{acknowledgments}
M.J. Rebou\c{c}as acknowledges the support of FAPERJ under a CNE E-26/102.328/2013 grant.
M.J.R. also thanks Glenn Starkman for important discussions about the circles-in-the-sky
searches, and CNPq for the grant under which this work was carried out.
G.I.G. thanks the CBPF for kind hospitality during the first part of this work.
We are also grateful to A.F.F. Teixeira for reading the manuscript and indicating some
omissions and misprints.
\end{acknowledgments}

\appendix
\section{} \label{Ap:Tangent}
\section{Tangent vectors}   

In this appendix we compute the tangent vector at a point ${\bf u}$ on the $3$-sphere
$\mathbb{S}^3$ of the geodesic joining ${\bf u}$ and another point ${\bf v}$ at a
distance $\mu$.

Let ${\bf u}$ and ${\bf p}$ be two points in the 3--sphere $\mathbb{S}^3$ such that
the distance between them is $\pi/2$. As shown in Ref.~\cite{Ratcliffe} (see Th. 2.1.5)
the geodesic line from ${\bf u}$ to ${\bf p}$ is
\begin{equation}
\label{OrtogGeodesic}
{\bf n}(s) = {\bf u} \cos s + {\bf p} \sin s \; .
\end{equation}
Since
\[
\langle {\bf n},{\bf u} \rangle = \cos s\; ,
\]
the geodesic is parametrized by the arc length $s$.

Let ${\bf v}$ be the point in this geodesic line located between ${\bf u}$ and ${\bf p}$
and at a distance $\mu$ from ${\bf u}$; then we have
\begin{equation}
{\bf v} = {\bf u} \cos \mu + {\bf p} \sin \mu \; .
\end{equation}
We can use this expression to write the geodesic in terms of ${\bf u}$ and ${\bf v}$,
in fact, we have
\begin{equation}
{\bf p} = \frac{1}{\sin \mu} ({\bf v} - {\bf u} \cos \mu) \; .
\end{equation}
Substituting this expression into~(\ref{OrtogGeodesic}) we obtain
\begin{equation}
{\bf n}(s) = \frac{1}{\sin \mu} \left[ {\bf u} \sin (\mu - s) +
{\bf v} \sin s \right] \; .
\end{equation}

Taking the derivative at $s=0$ one gets the tangent vector pointing from
${\bf u}$ to ${\bf v}$,
\begin{equation}
\label{TangentVec}
{\bf w} = \frac{1}{\sin \mu} \left( {\bf v} - {\bf u} \cos \mu \right) \; .
\end{equation}

\section{} \label{Ap:Theta}
\section{Deviation from antipodality $\theta$}

In this appendix we present details of the calculations of the deviation from
antipodality $\theta$ given by equation~(\ref{AntipodDist}).

Direct use of Eq.~(\ref{CircTangentVec}) and the fact that
\begin{equation}
\langle \mathbf{u},\mathbf{v}_1^{} \rangle = \langle \mathbf{u},\mathbf{v}_2^{}
\rangle = \cos \mu
\end{equation}
immediately yield
\begin{equation}
\label{CTVInnProd}
\langle \mathbf{w}_1^{} , \mathbf{w}_2^{} \rangle = \frac{1}{\sin^2 \mu}
\left(\langle \mathbf{v}_1^{},\mathbf{v}_2^{} \rangle - \cos^2 \mu \right) \,.
\end{equation}
Now,  since $\mathbf{v}_1^{} = \gamma \mathbf{u}$ and $\mathbf{v}_2^{}
= \gamma^{-1}\mathbf{u}$, the very definition of isometry implies that $\langle
\mathbf{v}_2^{},\mathbf{v}_1^{} \rangle = \langle \mathbf{u},\gamma^2\mathbf{u}
\rangle$, and thus~(\ref{Eq:Isometry}) and~(\ref{Eq:ObsPosition}) imply that
\begin{equation}
\label{Pre2ImageDist}
\langle \mathbf{v}_1^{},\mathbf{v}_2^{} \rangle = \cos 2\alpha \cos^2\rho + \cos
2\beta \sin^2\rho \,.
\end{equation}
Moreover, from Eq.~(\ref{ImageDist}) it easily follows  that
\begin{equation}
\cos \alpha - \cos \beta = \frac{\cos \mu - \cos \beta}{\cos^2 \rho} =
\frac{\cos \alpha - \cos \mu}{\sin^2 \rho} \,,
\end{equation}
{} from which we obtain
\begin{equation}
\label{RhoMuRelations}
\cos^2\rho = \frac{\cos \mu - \cos \beta}{\cos \alpha - \cos \beta}
\quad\mbox{and}\quad
\sin^2\rho = \frac{\cos \alpha - \cos \mu}{\cos \alpha - \cos \beta} \,\,.
\end{equation}
Substituting these relations into Eq.~(\ref{Pre2ImageDist}) one easily obtains
\begin{equation}
\label{2ndImageDist}
\langle \mathbf{v}_1^{},\mathbf{v}_2^{} \rangle = 2(\cos \alpha + \cos
\beta)\cos \mu - (2\cos \alpha \cos \beta + 1) \,,
\end{equation}
which upon substitution into~(\ref{CTVInnProd}) yields immediately
equation~(\ref{AntipodDist}).

\section{} \label{Ap:CriticalMu}
\section{Critical distance $\mu_0^{}$}

In this appendix we present details of the calculations of equation~(\ref{CriticalMu})
that gives the critical distance $\mu_0$ which maximize the deviation from antipodality.

Making the substitutions
\begin{equation}
\label{Subst}
A = \cos \alpha \qquad\mbox{,}\qquad B = \cos \beta \qquad\mbox{and}\qquad
x = \cos \mu
\end{equation}
into Eq.(\ref{AntipodDist}) it follows that, to minimize $\cos \theta$, i.e.
maximize $\theta$, one has to maximize the function
\begin{equation}
f(x) = \frac{(x-A)(x-B)}{x^2-1} \; .
\end{equation}
Since  $\alpha$ and $\beta$ are never supplementary angles, i.e. they never
sum up $\pi$, we in fact have $A+B\neq 0$.

Taking the first derivative $f'(x)$ and equating the numerator to zero one
obtains
\begin{equation}
\label{Poly}
x^2 - 2\ell x + 1 = 0 \; ,
\end{equation}
where
\begin{equation}
\label{ell}
\ell = \frac{AB+1}{A+B} \; .
\end{equation}
Equation~(\ref{Poly}) has real roots only if $\ell^2 \geq 1$, which holds
since $-1 < A < 1$ and $-1 < B < 1$. In fact, suppose first that $A+B>0$,
since $(1-A)(1-B) > 0$, then $\ell > 1$. On the other hand, suppose that
$A+B<0$, since $(1+A)(1+B) > 0$, then $\ell < -1$. In either case the
condition $\ell^2 \geq 1$ is fulfilled.

Moreover, since $x \leq 1$, the solution to Eq.~(\ref{Poly}) we are looking
for is
\begin{equation}
x_0^{} = \ell - \sqrt{\ell^2-1} \; ,
\end{equation}
so, using back the substitutions (\ref{Subst}) and (\ref{ell}), one gets
\begin{equation}
\label{CriticalMuP}
\cos \mu_0^{} = \frac{\cos \alpha \cos \beta + 1}{\cos \alpha + \cos \beta} -
\sqrt{\left(\frac{\cos \alpha \cos \beta + 1}{\cos \alpha +
\cos \beta}\right)^2 - 1} \; .
\end{equation}
To simplify this expression observe that
\begin{equation}
\label{CriticalMuQ}
(\cos \alpha \cos \beta + 1)^2 - (\cos \alpha + \cos \beta)^2 = \sin^2 \alpha
\sin^2 \beta \,,
\end{equation}
which, substituted into (\ref{CriticalMuP}) yields immediately~(\ref{CriticalMu}).

\section{} \label{Ap:Tables}
\section{Construction of examples}

In this appendix we give full details of the calculations involved in
the construction of concrete examples presented in Section~\ref{S:Examples}.

From (\ref{CriticalMuSin}) and $\delta - \epsilon = \alpha = \frac{2\pi}{p}$
we obtain
\begin{equation}
\sin \frac{2\pi}{p} = \left(\cos \frac{2\pi}{p} - \cos \mu_0^{}\right)
\tan \delta \; .
\end{equation}
To look for models with maximal deviation from antipodality, $\theta_{\mbox{\small max}} =
\pi - 2\delta$, larger than a threshold $\theta_0^{}$ it is enough to require
\begin{equation}
\label{IneqDeltaZero}
\delta < \frac{\pi - \theta_0^{}}{2} = \delta_0^{} \; ,
\end{equation}
and thus, $\tan \delta < \tan \delta_0^{}$. Recall now that in order to have
observable matched circles we need $\cos \mu_0^{} > \cos 2\chi_{\rm{LSS}}$; thus
using the fact that $\cos \frac{2\pi}{p} < 1$, together with the approximation
$\sin \frac{2\pi}{p} \approx \frac{2\pi}{p}$, since $p \gg 1$, one obtains the
inequality  
\begin{equation}
p > \frac{2\pi}{\left(1 - \cos 2\chi_{\rm{LSS}}^{}\right) \tan \delta_0^{}} \; .
\end{equation}

Taking the threshold $\theta_0^{} = 170^\circ$, i.e. $\delta_0^{} =
\frac{\pi}{36}$ in radians, for $\chi_{\rm{LSS}}^{} = 0.038$ one obtains the
lower bound $p > 24879$. However, since we are restricting our analysis to prime
values for $p$, this implies that $p \geq 24889$. This lower bound is very
tight, since numerical calculations show that actually $p \geq 24907$, as
shown in Tables \ref{Tb:Antipod1} and \ref{Tb:Antipod3}, which show examples
of lens spaces with relatively small values of the parameter $p$ whose canonical
generators have detectable maximal deviations from antipodality larger than  
$\theta_0^{} = 170^\circ$. Let us explain how these tables were built.
We note that Tables~\ref{Tb:Antipod2} and~\ref{Tb:Antipod4} are similarly constructed but correspond
to a threshold $\theta_0^{} = 90^\circ$, i.e. $\delta_0= \pi / 4$ radians. Let us explain
how these tables were constructed by referring to Tables~\ref{Tb:Antipod1} and~\ref{Tb:Antipod3}

In order to estimate efficiently the values for the parameter $q$ in Tables
\ref{Tb:Antipod1} and \ref{Tb:Antipod3} we recall the fact that all permissible
values for $q$ are given by $q_n^{} =q_0^{} - n$, for $n=1,2,3,\ldots, q_0^{}-2$,
where $q_0^{} = \frac{p-1}{2}$. Using equations~(\ref{AnglesN}) and~(\ref{IneqDeltaZero}),
for a maximal deviation from antipodality $\theta_{\mbox{\small max}}^{} > \theta_0^{}$ one needs
\begin{equation}
n < \frac{1}{2} \left[\left(1 - \frac{\theta_0^{}}{\pi}\right)p - 3\right] \; .
\end{equation}
For the threshold $\theta_0^{} = 170^\circ$ one obtains
\begin{equation}
n < \frac{1}{2} \left(\frac{p}{18} - 3\right) \; ,
\end{equation}
which implies
\begin{equation}
\label{IneqQn}
q_n^{} > \frac{17p}{36} + 1 \; .
\end{equation}
The second column in Table \ref{Tb:Antipod3} is calculated using this inequality.

\begin{table}[ht!]
\begin{tabular}{*4{c}}  
\hline  \hline
\ \ \; $p$ \; \ \ & \ \ \; $q$ \; \ \ \; & \; \ \ \; $\mu_0^{}$ \; \ \ & \ \ \; $\lambda$ \; \ \ \\
\hline
24\,907 & 11\,763 &  0.075978  &  $1.4^\circ$ \\
\hline
24\,917 & 11\,768 &  0.075978  &  $1.4^\circ$ \\
\hline
24\,919 & 11\,769 &  0.075978  &  $1.4^\circ$ \\
\hline
24\,923 & 11\,771 &  0.075978  &  $1.4^\circ$ \\
\hline
24\,943 & 11\,780 &  0.075923  &  $2.6^\circ$ \\
      & 11\,781 &  0.075979  &  $1.4^\circ$ \\
\hline
24\,953 & 11\,785 &  0.075923  &  $2.6^\circ$ \\
      & 11\,786 &  0.075979  &  $1.4^\circ$ \\
\hline
24\,967 & 11\,791 &  0.075868  &  $3.4^\circ$ \\
      & 11\,792 &  0.075923  &  $2.6^\circ$ \\
      & 11\,793 &  0.075979  &  $1.4^\circ$ \\
\hline
24\,971 & 11\,793 &  0.075868  &  $3.4^\circ$ \\
      & 11\,794 &  0.075924  &  $2.6^\circ$ \\
      & 11\,795 &  0.075979  &  $1.4^\circ$ \\
\hline
24\,977 & 11\,796 &  0.075868  &  $3.4^\circ$ \\
      & 11\,797 &  0.075924  &  $2.6^\circ$ \\
      & 11\,798 &  0.075979  &  $1.4^\circ$ \\
\hline\hline
\end{tabular}
\caption{Globally inhomogeneous lens spaces $L(p,q)$, with $p$ prime, spatial topology of nearly
flat universes with $\chi_{\rm{LSS}}^{} = 0.038$, for which both the maximal deviation
of antipodality, $\theta_{\mbox{\small max}}^{}$, and the radius of the matching
circles, $\lambda$, fall outside the range covered in the searches carried out
so far.
For each value of $p$ we display all the lens spaces $L(p,q)$ with the parameter
$q$ in the interval $[q^\bot,q^\top]$, where $q^\bot$ is the least value of $q$
for which $\theta_{\mbox{\small max}} > 170^\circ$, and $q^\top$ is the largest value of $q$
for which $\mu_0^{} < \chi_{\rm{LSS}}^{}$.
In  these universes $\theta_{\mbox{\small max}}^{} \approx 170^\circ$ for the
observer at $\rho_0\,$ [cf. Eq.~(\ref{CriticalMuSin})]. \label{Tb:Antipod3}} 

\end{table}

Recalling that $\theta_{\mbox{\small max}}^{} = \pi - 2 \delta$ and using (\ref{Eq:EpsDelta})
one obtains a simple expression to compute the maximal deviation from
antipodality in terms of the parameters $p$ and $q$ of the lens space,
\begin{equation}
\label{ThetaLens}
\theta_{\mbox{\small max}}^{} = \frac{2(q-1)\pi}{p} \,.
\end{equation}
Note that for each value of $p$ there is a least value of $q$ for which inequality
(\ref{IneqQn}) holds, let $q^\bot$ be this least value. From (\ref{ThetaLens}) one
can see that for $q > q^\bot$ the maximal deviation of antipodality is
$\theta_{\mbox{\small max}}^{} > \theta_{\mbox{\small max}}^\bot > \theta_0^{}$. For $q < q^\bot$, the maximal
deviation from antipodality is $\theta_{\mbox{\small max}}^{} < \theta_0^{} < \theta_{\mbox{\small max}}^\bot$.
The precise value of $\theta_{\mbox{\small max}}^{}$ is, however, very insensitive to the
parameters $p$ and $q$; in fact, a direct calculation shows that the maximal
deviation from antipodality for the spaces shown in Table \ref{Tb:Antipod3} ranges
between $\theta_{\mbox{\small max}}^{} = 170.005^\circ$ and $\theta_{\mbox{\small max}}^{} = 170.033^\circ$.

On the other hand, using equations~(\ref{CriticalMuSin}) and~(\ref{Eq:EpsDelta})
to write $\cos \mu_0^{}$ in terms of the parameters $p$ and $q$, it is
straightforward to check that the distance $\mu_0^{}$ is an increasing function
of $q$. So let $q^\top$ be the largest value of $q$ for which $\mu_0^{} < 2
\chi_{\rm{LSS}}^{}$; i.e., for which the generator of $L(p,q)$ has detectable 
maximal deviation of antipodality. For $q > q^\top$ one has $\mu_0^{} > 2
\chi_{\rm{LSS}}^{}$, so that the maximal deviation from antipodality is not
detectable. For $q < q^\top$, the maximal deviation from antipodality is detectable, 
however $\theta_{\mbox{\small max}}^{} < \theta_{\mbox{\small max}}^\top$.

\begin{table}[ht!]
\begin{tabular}{*4{c}}  
\hline  \hline
\ \ \; $p$ \; \ \ & \ \ \; $q$ \; \ \ \; & \; \ \ \; $\mu_0^{}$ \; \ \ & \ \ \; $\lambda$ \; \ \ \\
\hline
2\,203 & 552 &  0.075625  &  $5.7^\circ$ \\
     & 553 &  0.075733  &  $4.8^\circ$ \\
     & 554 &  0.075841  &  $3.7^\circ$ \\
     & 555 &  0.075949  &  $2.1^\circ$ \\
\hline
2\,207 & 553 &  0.075556  &  $6.2^\circ$ \\
     & 554 &  0.075664  &  $5.4^\circ$ \\
     & 555 &  0.075771  &  $4.4^\circ$ \\
     & 556 &  0.075879  &  $3.2^\circ$ \\
     & 557 &  0.075987  &  $2.1^\circ$ \\
\hline
2\,213 & 555 &  0.075507  &  $6.5^\circ$ \\
     & 556 &  0.075614  &  $5.8^\circ$ \\
     & 557 &  0.075721  &  $4.9^\circ$ \\
     & 558 &  0.075829  &  $3.9^\circ$ \\
     & 559 &  0.075937  &  $2.3^\circ$ \\
\hline\hline
\end{tabular}
\caption{Globally inhomogeneous lens spaces $L(p,q)$, with $p$ prime,
spatial topology of nearly flat universes with $\chi_{\rm{LSS}}^{} = 0.038$,
for which the radii of the matching circles, $\lambda$, fall outside the range
covered in the searches so far undertaken with either WMAP or Planck
data.
For each value of $p$ we display all the lens spaces $L(p,q)$ with
the parameter $q$ in the interval $[q^\bot,q^\top]$, where $q^\bot$ is the least
value of $q$ for which $\theta_{\mbox{\small max}} > 90^\circ$, and $q^\top$ is
the largest value of $q$ for which $\mu_0^{} < \chi_{\rm{LSS}}^{}$.
In  these universes $\theta_{\mbox{\small max}}^{} \gtrsim 90^\circ$
for the observer at $\rho_0\,$. Note that the values of the parameters
$p$ and $q$ are considerably smaller than those in Table~\ref{Tb:Antipod3},
and that the lengths of the intervals $[q^\bot,q^\top]$ are correspondingly
larger. \label{Tb:Antipod4}  }
\end{table}

We observe numerically that the size of the interval $[q^\bot,q^\top]$ increases
as $p$ increases. As one can see in Table \ref{Tb:Antipod3}, for values of $p$
from 24907 to 24923 one has $q^\bot = q^\top$, so there is only one lens space
with detectable maximal deviation from antipodality larger than $\theta_0^{} =
170^\circ$. For $p=24943$ and $p=24953$ it holds $q^\top = q^\bot + 1$, so for
each of these values of $p$ there are only two lens spaces with detectable maximal 
deviation from antipodality larger than $\theta_0^{} = 170^\circ$. For the five
primes between (and including) $p=24967$ and $p=24989$ it holds $q^\top = q^\bot
+ 2$, so there are three such spaces; in Table \ref{Tb:Antipod3} we show only those
corresponding to the first three values of $p$. Note that, according to our
analysis, larger values of $q$ correspond to smaller values of $\lambda$. In Table
\ref{Tb:Antipod1} we show the lens spaces $L(p,q^\bot)$ corresponding to the first
values of $p$ for which the size of the interval $[q^\bot,q^\top]$ increases in
one unit.

\newpage


\end{document}